\documentclass[11pt,showpacs,tightenlines,superscriptaddress]{revtex4}   
\pdfoutput=1

\usepackage{graphicx}
\usepackage{natbib}

\usepackage[T1]{fontenc}

\begin{document}
\title{High nitrogen-vacancy density diamonds for magnetometry applications}

 \author{V.~M.~Acosta}
    \address{
     Department of Physics, University of California,
     Berkeley, CA 94720-7300
    }
 \author{E.~Bauch}
    \address{
     Department of Physics, University of California,
     Berkeley, CA 94720-7300
    }
    \address{
     Technische Universit\"at Berlin,
     Hardenbergstra\ss e 28, 10623 Berlin, Germany
    }
 \author{M.~P.~Ledbetter}
    \address{
     Department of Physics, University of California,
     Berkeley, CA 94720-7300
    }
 \author{C.~Santori}
    \address{
     Hewlett-Packard Laboratories,
     1501 Page Mill Rd., Palo Alto, CA 94304
    }
 \author{K.-M.~C.~Fu}
     \address{
     Hewlett-Packard Laboratories,
     1501 Page Mill Rd., Palo Alto, CA 94304
    }
 \author{P.~E.~Barclay}
    \address{
     Hewlett-Packard Laboratories,
     1501 Page Mill Rd., Palo Alto, CA 94304
    }
 \author{R.~G.~Beausoleil}
     \address{
     Hewlett-Packard Laboratories,
     1501 Page Mill Rd., Palo Alto, CA 94304
    }
 \author{H.~Linget}
    \address{
     Ecole Normale Sup\'erieure de Cachan,
     61 Avenue du Pr\'esident Wilson, 94235 Cachan CEDEX, France
    }
 \author{ J.~F.~Roch}
    \address{
     Laboratoire de Photonique Quantique et Mol\'eculaire (CNRS UMR 8537),
     Ecole Normale Sup\'erieure de Cachan,
     61 Avenue du Pr\'esident Wilson, 94235 Cachan CEDEX, France
    }
 \author{F.~Treussart}
    \address{
     Laboratoire de Photonique Quantique et Mol\'eculaire (CNRS UMR 8537),
     Ecole Normale Sup\'erieure de Cachan,
     61 Avenue du Pr\'esident Wilson, 94235 Cachan CEDEX, France
    }
 \author{S.~Chemerisov}
    \address{
     Argonne National Laboratory,
     Argonne, IL, 60439, U.S.A.
    }
 \author{W.~Gawlik}
    \address{
     Center for Magneto-Optical Research,
     Institute of Physics, Jagiellonian University,
     Reymonta 4, 30-059 Krak\'ow, Poland
    }
 \author{D.~Budker}
    \address{
     Department of Physics, University of California,
     Berkeley, CA 94720-7300
    }
    \address{Nuclear Science Division, Lawrence Berkeley National Laboratory,
     Berkeley CA 94720, USA
    }
    \address{budker@berkeley.edu}

\date{\today}

\begin{abstract}
Nitrogen-vacancy (NV) centers in millimeter-scale diamond samples were produced by irradiation and subsequent annealing under varied conditions. The optical and spin relaxation properties of these samples were characterized using confocal microscopy, visible and infrared absorption, and optically detected magnetic resonance. The sample with the highest NV$^{\mbox{-}}$ concentration, approximately $16~{\rm ppm}$ ($2.8\times10^{18}~{\rm cm^{-3}}$), was prepared with no observable traces of neutrally-charged vacancy defects. The effective transverse spin-relaxation time for this sample was $T_2^{\ast}=118(48)~{\rm ns}$, predominately limited by residual paramagnetic nitrogen which was determined to have a concentration of $49(7)~{\rm ppm}$. Under ideal conditions, the shot-noise limited sensitivity is projected to be $\sim 150~{\rm fT/\sqrt{Hz}}$ for a $100~{\rm \mu m}$-scale magnetometer based on this sample. Other samples with NV$^{\mbox{-}}$ concentrations from $.007~{\rm to}~12~{\rm ppm}$ and effective relaxation times ranging from $27$ to over $291~{\rm ns}$ were prepared and characterized.
\end{abstract}
\maketitle

\section{Introduction}
Optical magnetometers based on alkali vapor cells can measure magnetic fields with exceptional sensitivity and without cryogenics; however spin-altering collisions limit the sensitivity of small sensors \cite{BUD2007}. Paramagnetic impurities in diamond, on the other hand, are a promising system for millimeter- and micrometer-scale magnetometers, because they exhibit long spin coherence times (several ms \cite{BAL2009}) over temperatures ranging from liquid helium to well above room temperature. Nitrogen-vacancy (NV) centers are particularly promising because they have a spin-triplet ground state and convenient optical transitions, allowing for efficient optical pumping and magnetic sensitivity. Recently, single NV-centers were used for nanometer-scale magnetometry \cite{MAZ2008NATURE,BAL2008} and magnetometry based on high-density ensembles of NV$^{\mbox{-}}$ centers was proposed \cite{TAY2008}. Here we discuss preparation of diamond samples with parameters optimized for room-temperature, optical NV-ensemble magnetometers.

The parameters for sample preparation most relevant to magnetometry are the concentration and homogeneity of the NV$^{\mbox{-}}$ centers, as well as the abundance of other impurities such as neutral nitrogen-vacancy centers (NV$^0$) and substitutional nitrogen in the lattice. There is a wealth of techniques for growth, irradiation, and annealing of diamonds that can be considered (see, for example, Refs. \cite{KAN2000,BAC1993,PAL1994,WAL2007} and references therein). In this work, we prepared ten samples under different conditions and observed the optical properties at each stage of development using confocal microscopy, visible and infrared absorption, and optically detected magnetic resonance (ODMR). The results form an initial matrix that will help identify the most important variables and pinpoint optimal ranges in the sample preparation process.

\section{Experimental Procedure}
\subsection{Sample parameters and irradiation}
\label{Sec_irradiation}
\begin{table}[ht]

\centering
    \begin{tabular}{c c c c c c c}
      \hline
      \hline
      \# & ~Company~ & ~Synthesis~ & ~[N]~(ppm)~ & ~Radiation~ & ~Dose~(cm$^{-2}$)~ & ~Anneal Temp.~[1st,2nd,3rd]~($^\circ$C)\\
      \hline
      1 & Sumitomo & HPHT & $\lesssim100$ & e$^{\mbox{-}}$ & $4.0 \times 10^{18}$ & [700,875,1050]\\ 
      2 & Sumitomo & HPHT & $\lesssim100$ & e$^{\mbox{-}}$ & $9.8 \times 10^{18}$ & \textquotedbl \\ 
      3 & Element-6 & CVD & $\lesssim1$ & e$^{\mbox{-}}$ & $4.0 \times 10^{17}$ & \textquotedbl \\ 
      4 & Element-6 & CVD & $\lesssim1$ & e$^{\mbox{-}}$ & $8.0 \times 10^{16}$ & \textquotedbl \\ 
      5 & Element-6 & HPHT & $\lesssim200$ & e$^{\mbox{-}}$ & $8.2 \times 10^{18}$ & \textquotedbl \\ 
      6 & Element-6 & HPHT & $\lesssim200$ & e$^{\mbox{-}}$ & $9.8 \times 10^{18}$ & \textquotedbl \\ 
      7 & Sumitomo & HPHT & $\lesssim100$ & e$^{\mbox{-}}$ & $9.8 \times 10^{18}$ & \textquotedbl \\ 
      8 & Element-6 & HPHT & $\lesssim100$ & e$^{\mbox{-}}$ & $4.0 \times 10^{17}$ & \textquotedbl \\ 
      9 & Element-6 & HPHT & $\lesssim100$ & $H^{+}$ & $1.0 \times 10^{16}$ & [500,875,1050]\\ 
      10 & Element-6 & HPHT & $\lesssim100$ & $H^{+}$ & $1.0 \times 10^{16}$ & [800,875,1050] \\ [1ex]
      \hline
    \end{tabular}
    \caption{Sample characteristics. Initial nitrogen concentration, [N], is given as specified by manufacturer. Each annealing was carried out for $2 - 2.5$ hours. \label{tab:prep}}
\end{table}

Ten diamonds were fabricated commercially by Sumitomo and Element-6 using either chemical-vapor deposition (CVD) or high-pressure, high-temperature (HPHT) synthesis. The samples are millimeter-sized diamond crystals with an initial concentration of nitrogen impurities ranging from less than $1~{\rm ppm}$ to $200~{\rm ppm}$ ($1~{\rm ppm}$ corrsponds to $1.76\times10^{17}~{\rm cm^{-3}}$ in diamond). In the case of the HPHT samples, growth sectors with high and low nitrogen concentrations were identified and three of these sectors in Sample 5 were studied separately (see Sec. \ref{sec:odmr}). The crystals were subjected to either proton or electron irradiation at varying doses. Electron irradiation was done at Argonne National Laboratory using 3.0-MeV electrons (e$^{\mbox{-}}$) from a Van de Graaff generator. Thermal contact with a water-cooled copper target was used to keep the temperature of the samples below $150^{\circ}$C during the irradiation process in order to reduce vacancy-carbon recombinations \cite{NEW2002}. The irradiation doses were chosen so that the resulting vacancy concentrations would be within two orders of magnitude of the initial nitrogen concentration, as discussed in Sec. \ref{sec:ntonv}. The proton irradiation was performed at the Conditions Extrêmes et Matériaux: Haute Température et Irradiation (CEMHTI) laboratory using 2.4-MeV protons (H$^{+}$), and the samples were maintained below $80^{\circ}$C during irradiation. Table \ref{tab:prep} shows the specific conditions under which each sample was prepared.

\subsection{Optical characterization and annealing}

After the irradiation, optical characterization of each electron-irradiated sample was performed using confocal-microscopy at room temperature \cite{SAN2006OPTEX}. Light from a commercial solid-state $532\mbox{-}{\rm nm}$ laser was spatially filtered using a single-mode fiber, collimated, spectrally filtered, and combined into the collection path by a dichroic beamsplitter, exciting an optical transition associated with the NV$^{\mbox{-}}$ center's $^3E$ excited-state manifold \cite{MAN2006}. The excitation beam was focused to a sub-micron spot using a 0.6-numerical-aperture microscope objective with 40x magnification. The resulting photoluminescence (PL) was collected through the same objective, spectrally filtered to remove reflected laser light, imaged onto a pinhole, and analyzed. For spatial imaging, the microscope objective was mounted on a closed-loop triple-axis piezoelectric translation stage, and the collected photoluminescence was detected using a silicon avalanche photodiode, with filters used to select the NV$^{\mbox{-}}$ phonon sidebands (PSB) in the spectral range $\sim 647\mbox{-}804$ nm. For spectroscopy, the collected photoluminescence was sent to a spectrometer incorporating a cooled CCD detector, where the spectrum was recorded for various locations throughout each sample. The NV$^{0}$ and NV$^{\mbox{-}}$ zero-phonon lines (ZPL) were fit to Gaussian lineshapes after subtracting the local linear background. Imaging tests on single NV centers in an Electronic Grade ultra-pure CVD diamond purchased from Element-Six indicated an effective measurement volume of $5.2~{\rm \mu m}^3$, and the single-center fluorescence spectrum provided the absolute scale for determining the NV$^{\mbox{-}}$ concentrations in the higher NV$^{\mbox{-}}$ density samples.  

Following irradiation and after the initial round of PL measurements, the eight electron-irradiated samples were subjected to thermal annealing at $700^{\circ}$C in the presence of a forming gas (0.96 atm Ar, 0.04 atm H$_{2}$), in order to prevent surface oxidation, for $2-2.5$ hours. The two proton-irradiated samples were annealed for two hours in vacuum, one sample at a temperature of $800^{\circ}$C (Sample $10$) and the other at $500^{\circ}$C (Sample $9$). Afterwards, each sample was analyzed with confocal spectroscopy at room temperature. This process was repeated for nine of the samples at an annealing temperature of $875^{\circ}$C and then $1050^{\circ}$C in the presence of the same forming gas. After the final annealing, a Varian Cary 50 UV-VIS spectrometer and Nicolet Magna IR-750 Fourier-transform infrared (FTIR) spectrometer were used to measure visible and infrared absorption spectra for some of the electron-irradiated samples. Additionally, confocal microscopy was used for ODMR measurements and the effective transverse spin-relaxation times were determined.

\section{Results}
\subsection{Proton-irradiated samples}
\label{sec:protons}
Development of micro-magnetometers in certain geometries may benefit from the use of proton irradiation, where damage is confined to a smaller volume than with electron irradiation. In this section we present results of the optical characterization of two identical proton-irradiated samples (Samples 9 and 10).

\subsubsection{Stopping Range of Ions in Matter (SRIM) and spectroscopic measurements}
\label{sec:srim}
When charged particles traversing a medium come near to rest, the stopping power dramatically increases, causing a thin layer of radiation damage to develop. This effect is pronounced for non-relativistic ions in diamond. The ionization losses limit penetration to a thin layer near the surface, where radiation-induced vacancies in the carbon lattice are formed. In contrast, relativistic electrons make it through the $0.5 - 1~{\rm mm}$ thick samples, and, consequently, the vacancies are distributed relatively evenly.
 \begin{figure}
    \includegraphics[width=.8\textwidth]{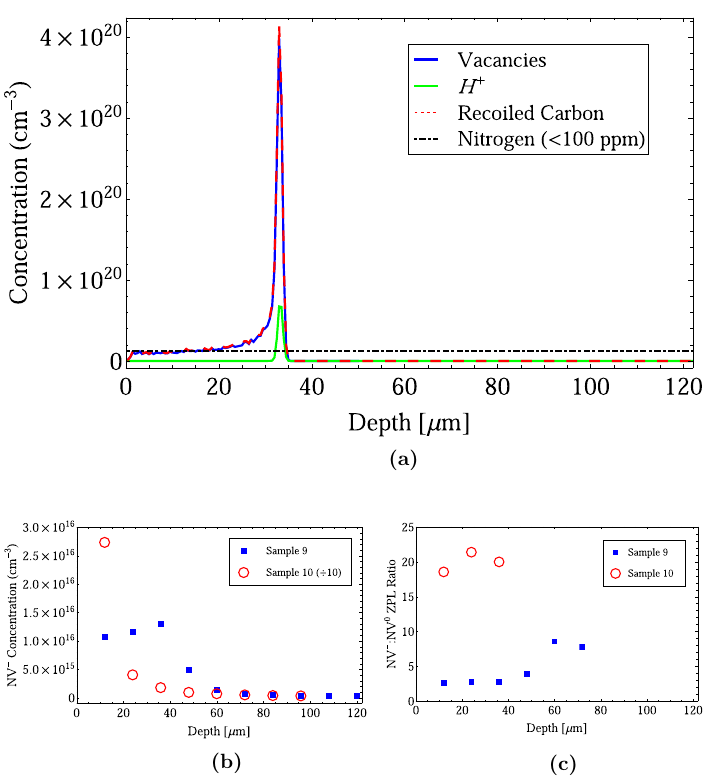}
    \caption{\label{fig:protonfig}(Color online) (a) Monte-carlo simulation of radiation damage as a function of penetration depth for 2.4-MeV proton radiation, using the SRIM software. (b) Depth profile of NV$^{\mbox{-}}$ centers, taken by normalizing the integrated ZPL intensity by that of a single NV center, after the first annealing (see text). The concentrations for Sample $10$ have been divided by a factor of 10 for visualization. Errors on the concentration estimates are discussed in Sec. \ref{sec:NVconcentrations}. (c) NV$^{\mbox{-}}$:NV$^0$ ZPL intensity ratio as a function of depth after the first annealing. The lightpower was $100~{\rm \mu W}$. For Sample 10, beyond a depth of $\sim36~{\rm \mu m}$, the NV$^0$ ZPL became indistinguishable from the background.}
\end{figure}

Figure \ref{fig:protonfig}(a) shows the results of proton-irradiation Monte-Carlo simulations made using the Stopping and Range of Ions in Matter (SRIM) software \cite{ZIE2008}. The diamond sample was modeled as a pure $^{12}$C layer with density $3.52~{\rm g/cm^3}$ and displacement energy $45~{\rm eV}$ \cite{BOU1976,KOI1992,SAA1998}. For the H$^+$ energy of 2.4 MeV, the simulations predict a dense layer of vacancies, recoiled interstitial carbon atoms, and stopped H$^+$ ions at a depth of $\sim 35~{\rm \mu m}$, and almost no radiation damage deeper into the sample. If vacancy-to-NV conversion is relatively even throughout the sample, then one might expect a similar depth profile of NV concentration.

Figure \ref{fig:protonfig}(b) shows the estimated NV$^{\mbox{-}}$ concentration, taken by normalizing the integrated ZPL intensity by that of a single NV center, as a function of depth for the proton-irradiated samples after the first annealing. Figure \ref{fig:protonfig}(c) displays the NV$^{\mbox{-}}$:NV$^0$ ZPL intensity ratio under the same conditions as Fig. \ref{fig:protonfig}(b) and is discussed later in Sec. \ref{sec:proton_discuss}. In Fig. \ref{fig:protonfig}(b), Sample $9$ (after annealing at $500^{\circ}$C) shows a thin layer of NV$^{\mbox{-}}$ emission with a small peak at $\sim35~{\rm \mu m}$, followed by a rapid decrease in concentration with depth. The peak intensity, however, is not nearly as dramatic as in the vacancy profile predicted by SRIM. Sample $10$ (after the first annealing at $800^{\circ}$C) shows a sharp, monotonic decrease throughout the first $35~{\rm \mu m}$ that is not at all consistent with the vacancy profile predicted by SRIM. Upon further measurements (see Fig. \ref{fig:decon}(a) in Sec. \ref{sec:pl}), it was seen that the qualitative features in the depth profiles of Sample $10$ varied greatly for different locations. It is worth mentioning that in this sample the NV$^{\mbox{-}}$ concentration varied by more than an order of magnitude for different locations all at a depth of $\sim 12~{\rm \mu m}$ from the surface. There was one location that had a concentration nearly three orders of magnitude lower than the mean. As shown in the next section, these spatial inhomogeneities in NV$^{\mbox{-}}$ concentration are related to the presence of multiple growth sectors, with widely varying nitrogen concentrations, in these HPHT-synthesized diamonds.

\subsubsection{Photoluminescence depth profiles}
\label{sec:pl}

\begin{figure}
    \includegraphics[width=.8\textwidth]{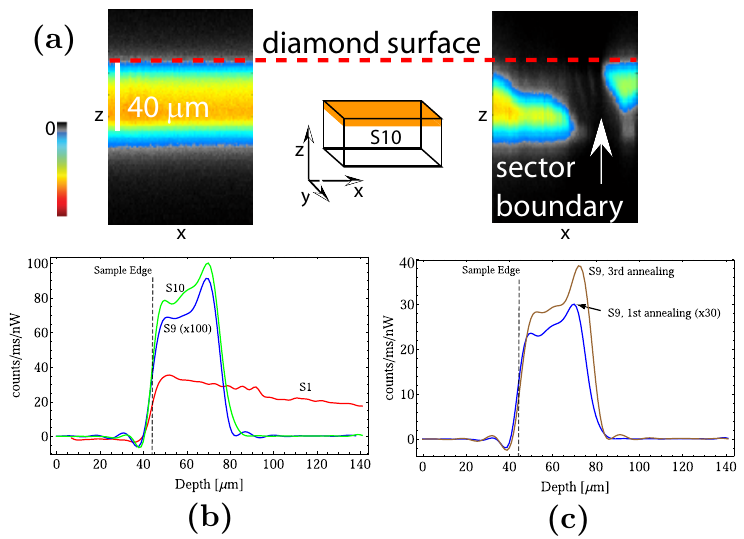}
    \caption{\label{fig:decon}(Color online) (a) Depth profiles of the phonon-sideband PL intensity for Sample 10 at a typical location (left) and at a sector boundary (right). (b) Deconvolved intensity as a function of depth for Samples $1,9,~{\rm and}~10$ (S1, S2, etc.) after the first annealing at $700$, $500$, and $800~^{\circ}$C, respectively. The PL intensity is normalized by the incident laser power ($5, 153,~{\rm and}~1.9~{\rm nW}$ for Samples $1,9,~{\rm and}~10$, respectively). The signal for Sample $9$ is multiplied by a factor of 100 for visualization. (c) PL deconvolution for Sample $9$ after the first and third annealing. The PL intensity after the first annealing has been multiplied by a factor of 30. The damped oscillations are artifacts of the signal processing caused by rapid changes in concentration.}
\end{figure}

The spectroscopic measurements were followed by spatially resolved PL measurements where the position of the focus was varied by translating the microscope objective with a closed-loop, triple-axis piezoelectric stage and the fluorescence, in the $647\mbox{-}804~{\rm nm}$ range, was collected. Figure \ref{fig:decon}(a) shows examples of raw data both for a typical scan (left side) and one around a sector boundary (right). After averaging over the lateral dimension, x, the resulting depth profile was deconvolved using a similarly-processed profile from a sample containing a very thin layer ($\lesssim200~{\rm nm}$) of NV centers as the response function, in conjunction with an optimized Wiener filter \cite{PRE1992}. Figure \ref{fig:decon}(b) shows the results for three different samples: both of the proton-irradiated samples as well as Sample 1, a high nitrogen-content electron-irradiated sample, for comparison. These profiles were taken after the final annealing at $1050^{\circ}$C and were post-selected to ensure that they were from a single growth-sector.  As expected for relativistic-electron irradiation, Sample $1$ has a relatively even depth profile throughout the sample. The gradual decrease in intensity with depth is consistent with absorption of the $532~{\rm nm}$ laser light, as seen in Sec. \ref{sec:absorption}. Samples $9$ and $10$, on the other hand, exhibit emission only from a thin layer of NV$^{\mbox{-}}$ centers near the surface. The edge of this layer is at approximately the same depth as the edge in the vacancy profile predicted by SRIM, $\sim 35~{\rm \mu m}$ from the surface, but the sharp peak at the end of the layer is absent. Figure \ref{fig:decon}(c) shows the depth profile in a single-sector region for Sample 9 after both the first and final annealing; the results are discussed below.

\subsubsection{Discussion of fluorescence depth profiles}
\label{sec:proton_discuss}
There could be several potential explanations for the deviation of the observed NV$^{\mbox{-}}$ fluorescence depth profiles from the vacancy distribution predicted by SRIM:

\begin{enumerate}
    \item The ZPL intensity and the integrated PSB fluorescence are poor indicators of the actual NV$^{\mbox{-}}$ concentration.
    \item The vacancies diffuse tens of ${\rm \mu m}$ during annealing.
    \item The concentration of nitrogen is insufficient to convert most of the vacancies to NV$^{\mbox{-}}$ centers at the stopping peak.
    \item Other products of the irradiation, either the interstitial carbon, stopped H$^+$ ions, or the vacancies themselves, inhibit NV$^{\mbox{-}}$ formation by forming stable defects with the vacancies.
\end{enumerate}

We now briefly discuss each of these possibilities and show that the first three explanations are not supported by our observations.

1. In the He$^+$ ion implantation work of Ref. \cite{WAL2007}, a similar deviation from the SRIM vacancy profiles was seen in the NV$^{\mbox{-}}$ ZPL intensities for various radiation doses. The primary cause posited in that work was radiation absorption by vacancies and graphitization defects. In order to set a limit on the absorption by such defects in our samples, we monitored the intensity of the diamond's first-order Raman peak at $573~{\rm nm}$ during the spectroscopic measurements whenever it was distinguishable from the background. In every instance, the $573~{\rm nm}$ peak intensity did not change by more than a factor of two over the first $\sim100~{\rm \mu m}$ of the sample, and at least some of this attenuation was due to absorption by the NV$^{\mbox{-}}$ centers themselves (see Sec. \ref{sec:absorption}). To our knowledge, there is no defect present in large concentrations in these samples that absorbs strongly in the NV$^{\mbox{-}}$ fluorescence region but not at $573~{\rm nm}$. The neutral vacancy center, GR1, for example, absorbs roughly equally in these two spectral regions \cite{COL1981,ALL1998}. Therefore a factor of approximately two is the maximum possible attenuation of the NV$^{\mbox{-}}$ fluorescence due to absorption by other defects. The $35\mbox{-}{\rm \mu m}$ peak height predicted by the SRIM vacancy depth profile is several orders of magnitude larger than the peak heights of the NV$^{\mbox{-}}$ depth profiles in Figs. \ref{fig:protonfig}(b) and \ref{fig:decon}(b), so radiation absorption cannot explain this deviation.

2. The results shown in Fig. \ref{fig:decon}(c) give evidence that vacancy diffusion cannot explain the deviation from the SRIM vacancy depth profile. The magnitude of NV$^{\mbox{-}}$ fluorescence in the depth profile for Sample 9 exhibits an increase of approximately a factor of 35 between measurements taken after the first and third annealing, signifying that vacancies are diffusing to form NV$^{\mbox{-}}$ centers during the second and/or third annealing. However, based on the shape of these depth profiles, and taking into account the artifacts introduced by the deconvolution, a limit of $\delta r \lesssim 1.5~{\rm \mu m}$ can be placed on the spatial range of diffusion of vacancies in this sample during annealing at $875^{\circ}$C and $1050^{\circ}$C for a time of $t=2~{\rm hours}$ each. This is not enough diffusion to account for the large difference in peak height predicted by the SRIM vacancy depth profile. Assuming the diffusion coefficient at $875^{\circ}$C is smaller than that at $1050^{\circ}$ and that the vacancies are not annihilated by other processes during annealing, this corresponds to an upper bound on the diffusion coefficient at $1050^{\circ}$C of $D_V\approx(\delta r)^2/(4 t)\lesssim4 \times 10^{-13}~{\rm cm^2/s}$.

3. To address the possibility that insufficient nitrogen levels limit the overall NV$^{\mbox{-}}$ concentration at the stopping peak, a rough measurement of the levels of substitutional nitrogen available for NV formation was made by monitoring the ZPL of the neutral NV$^0$ center at $575~{\rm nm}$. Substitutional nitrogen is a donor in diamond, believed to be responsible for the negative charge on the NV$^{\mbox{-}}$ center, and consequently a shortage of nitrogen leads to a greater proportion of neutral NV$^0$ centers \cite{MIT1996,WAL2007,SAN2009}. Figure \ref{fig:protonfig}(c) shows the NV$^{\mbox{-}}$:NV$^0$ ZPL intensity ratio for Samples $9$ and $10$ for the same depths as in Fig. \ref{fig:protonfig}(b). The high ratio for Sample 10 is inconsistent with donor depletion. The ratio for Sample $9$ is lower compared to that of similar samples irradiated at lower doses (${\rm NV^{\mbox{-}}:NV^0\approx 1 - 40}$ (see Fig. \ref{fig:anneal}(b)), but the absence of a sharp dip near $35~{\rm \mu m}$ suggests that electron depletion cannot account for the deviation from the SRIM vacancy depth profile.

4. The possibility that other radiation damage inhibits NV formation cannot be verified in this study, but it is not inconsistent with our observations. As seen in Figure \ref{fig:protonfig}(a), the recoiled carbon atoms are the most abundant defect present in the diamond, with more than $30$ times the nitrogen concentration. The deposited H$^+$ also comprise over five times the nitrogen concentration. At these levels, it is more likely that during annealing the diffusing vacancies will recombine, or form other color centers, with these interstitial atoms before they encounter a nitrogen atom. ESR measurements of the interstitial carbon color center, R2, suggest that some recombination occurs during irradiation unless the diamond is cooled well below room temperature \cite{HUN2000,NEW2002} (not the case here), and that during annealing the remaining R2 centers become mobile at a threshold temperature in the $400-700^{\circ}$C range {\cite{HUN2000,ALL1998,NEW2002}}. Samples $9~{\rm and}~10$ are actually two halves of the same diamond, but their first annealing was done at different temperatures: $500^{\circ}$C and $800^{\circ}$C, respectively. The temperature of these samples was maintained below $80^{\circ}$C during irradiation, yet their depth profiles after these different annealing temperatures show similar deviation from SRIM vacancy profiles. This suggests that, if interstitial recombination is the mechanism responsible for this deviation, the onset of self-interstitial mobility occurs at or below an annealing temperature of $500^{\circ}$C. This limit still falls within the $400-700^{\circ}$C range of values for the onset of self-interstitial mobility found in the literature.

In addition to recombination with interstitial carbon atoms, vacancies are also known to combine with hydrogen atoms to form VH$^{\mbox{-}}$ \cite{GLO2004} and NVH$^{\mbox{-}}$ \cite{GLO2003} defects. As we were not setup to observe ESR where these paramagnetic centers are best detected, they were not studied here, but they may limit NV formation and dephase NV$^{\mbox{-}}$ spins in magnetometry applications.

In summary, the proton-irradiation and annealing produced NV centers in a confined volume of depth $\sim35~{\rm \mu m}$. The depth profiles of the NV$^{\mbox{-}}$ center concentrations showed a significant deviation from the expected damage profiles. We considered possible causes of such a deviation and surmised that excess radiation damage likely inhibited NV formation. The resulting [NV$^{\mbox{-}}$] distributions may actually be better for some micro-magnetometer configurations, but more work is required to understand the details of the radiation damage.

\subsection{Electron-irradiated samples}
\label{Sec_electron}
In contrast to proton irradiation, irradiation with electrons leaves a relatively even distribution of damage throughout the sample, enabling the preparation of millimeter-sized samples for ultra-sensitive magnetometry. Here we present and discuss the optical and magnetic-resonance characterization of electron-irradiated samples (Samples 1-8) prepared under varied conditions.

\subsubsection{Fluorescence spectra and calibrations}
\label{sec:fluoresce}
\begin{figure}
    \includegraphics[width=.88\textwidth]{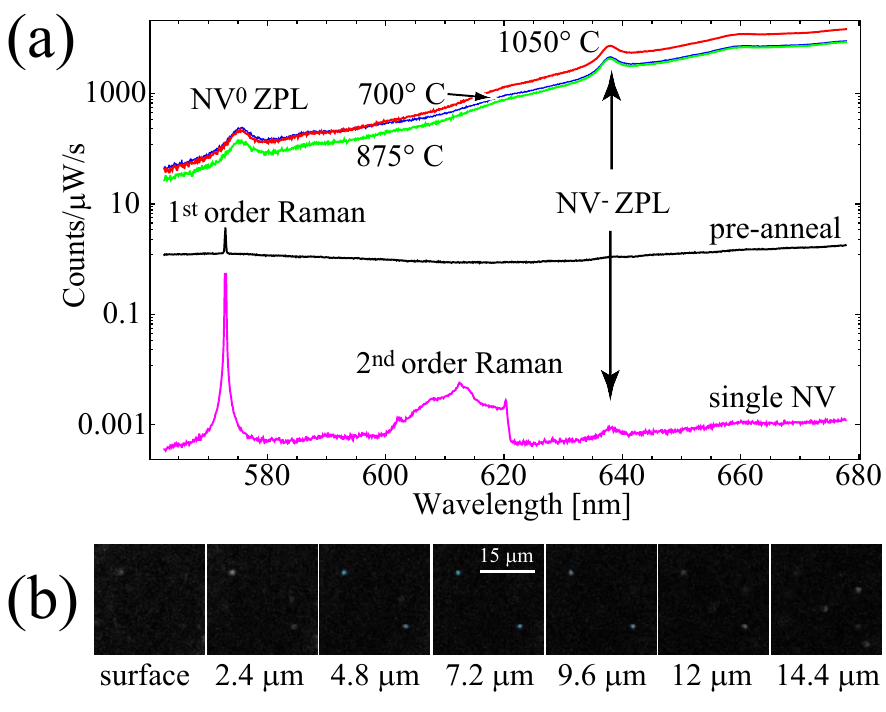}
    \caption{\label{fig:fluorescence}(Color online) Fluorescence spectrum at each stage of annealing for Sample $2$ (see Tab. \ref{tab:prep}), as well as the spectrum of a single NV center in the ultra-pure CVD diamond used for calibrations. The measurement volume is approximately $5.2~{\rm \mu m}^3$ located $\sim 12~{\rm \mu m}$ from the surface. The important spectral features are labeled (as described in the text). (b) Lateral (xy) scans of the ultra-pure CVD diamond showing fluorescence in the $647-804~{\rm nm}$ spectral region from two of the single NV centers used to determine the fluorescence collection volume. The range for all of the scans is $30~\rm{\mu m} \times 30~\rm{\mu m}$}.
\end{figure}

Figure \ref{fig:fluorescence}(a) shows the room-temperature $532\mbox{-}{\rm nm}$ laser-induced fluorescence spectrum for Sample $2$, a $2 \times 2 \times 1~{\rm mm}^3$ Sumitomo HPHT sample irradiated with electrons at a dose of $9.8 \times 10^{18}~{\rm cm}^{-2}$  (see Table \ref{tab:prep}), for each stage of annealing, as well as the spectrum for a single NV in a non-irradiated Electronic Grade ultra-pure CVD diamond. The collection volume was located near the surface of the sample in order to minimize the effects of laser and fluorescence absorption. Before annealing, the main spectral feature is a sharp peak at $573~{\rm nm}$ which is the first-order Raman peak of the pure diamond lattice.  Additionally, the broad second-order Raman feature is visible for the single-NV center in the $600-625~{\rm nm}$ region. After annealing, the much brighter NV$^{0}$ and NV$^{\mbox{-}}$ zero-phonon lines, at $575~{\rm nm}$ and $638~{\rm nm}$, respectively, with linewidths of approximately $1.5~{\rm nm}$ full width at half maximum, become prominent features in the spectrum. The monotonically increasing background is due to the overlapping broad phonon sidebands. For more details on the spectral features, see, for example, Ref. \cite{ZAI2001}.

In order to quantify the NV concentrations, the fluorescence collection volume was determined by taking several lateral scans of the ultra-pure CVD diamond at different depths and recording the $647-804~{\rm nm}$ fluorescence of single NVs, as illustrated in Fig. \ref{fig:fluorescence}(b). A collection volume of $\sim5.2~{\rm \mu m}^3$ was calculated from these scans. This value, together with the integrated ZPL intensity at $638~{\rm nm}$ from the spectrum of one of the single NV centers (Fig. \ref{fig:fluorescence}(a)), were used to determine the NV concentrations throughout this work. We note that concentrations obtained in this way are expected to be a lower bound, as the laser power used for these single NV scans was just above saturation, and the effects of PL quenching due to non-radiative energy transfer to nearby impurities \cite{DAV1973} were not quantified.  Nonetheless, as will be shown in Sec. \ref{sec:absorption} and \ref{sec:IRabsorption}, this method produced concentrations values that are in close agreement with absorption-based methods \cite{DAV1992,LAW1998}. Furthermore, a fluorescence-based approach enables the measurement of NV concentrations in optically thin samples.

\subsubsection{NV concentrations}
\label{sec:NVconcentrations}
\begin{figure}[ht]
    \begin{minipage}[b]{1\linewidth}
        \centering
        \includegraphics[width=.67\textwidth]{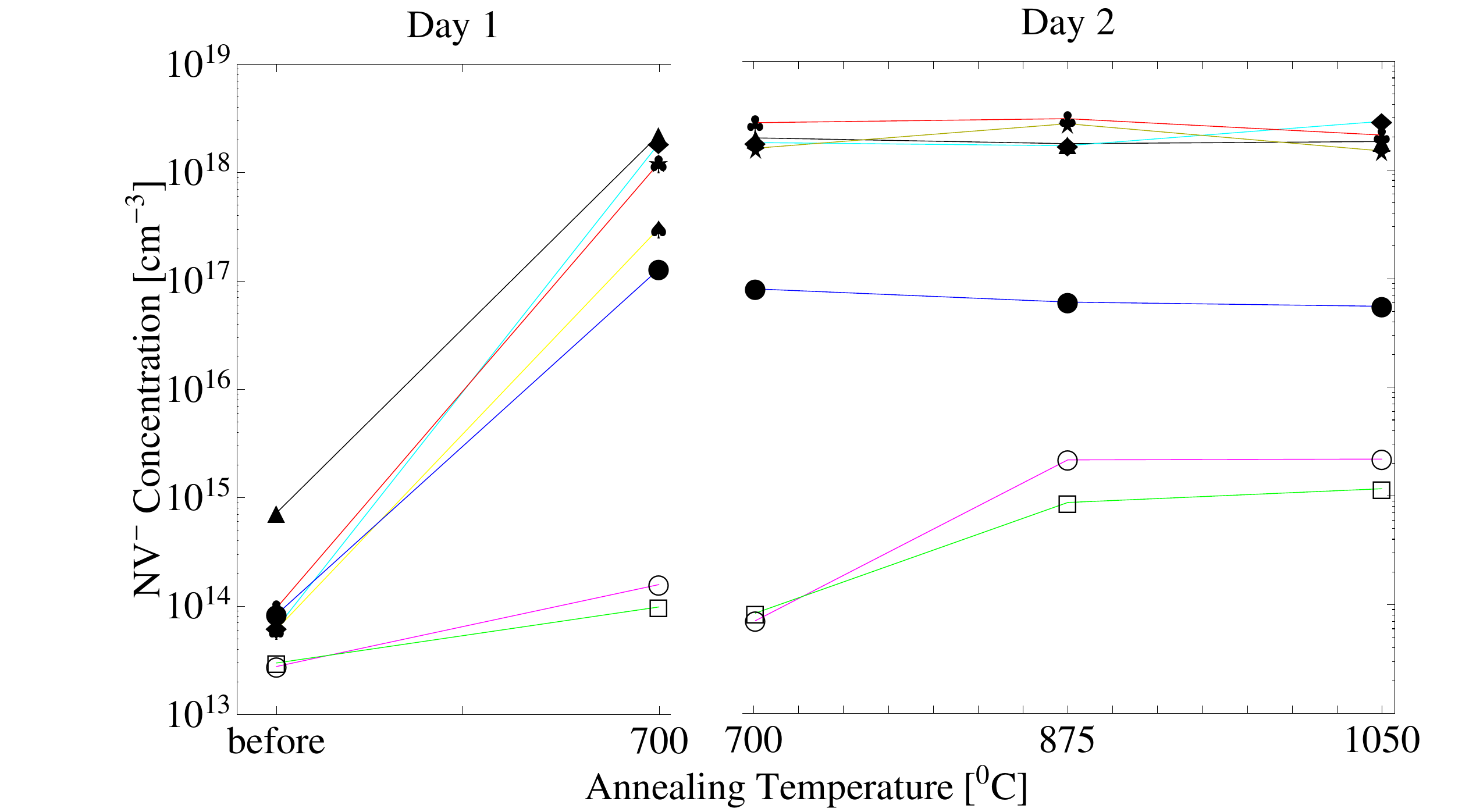}
        \vspace{1 cm}
        \begin{minipage}[b]{1\linewidth}
            \centering
            \textbf{(a)}
        \end{minipage}
        \vspace{-1.5 cm}
    \end{minipage}
    \begin{minipage}[b]{1\linewidth}
        \centering
        \includegraphics[width=.72\textwidth]{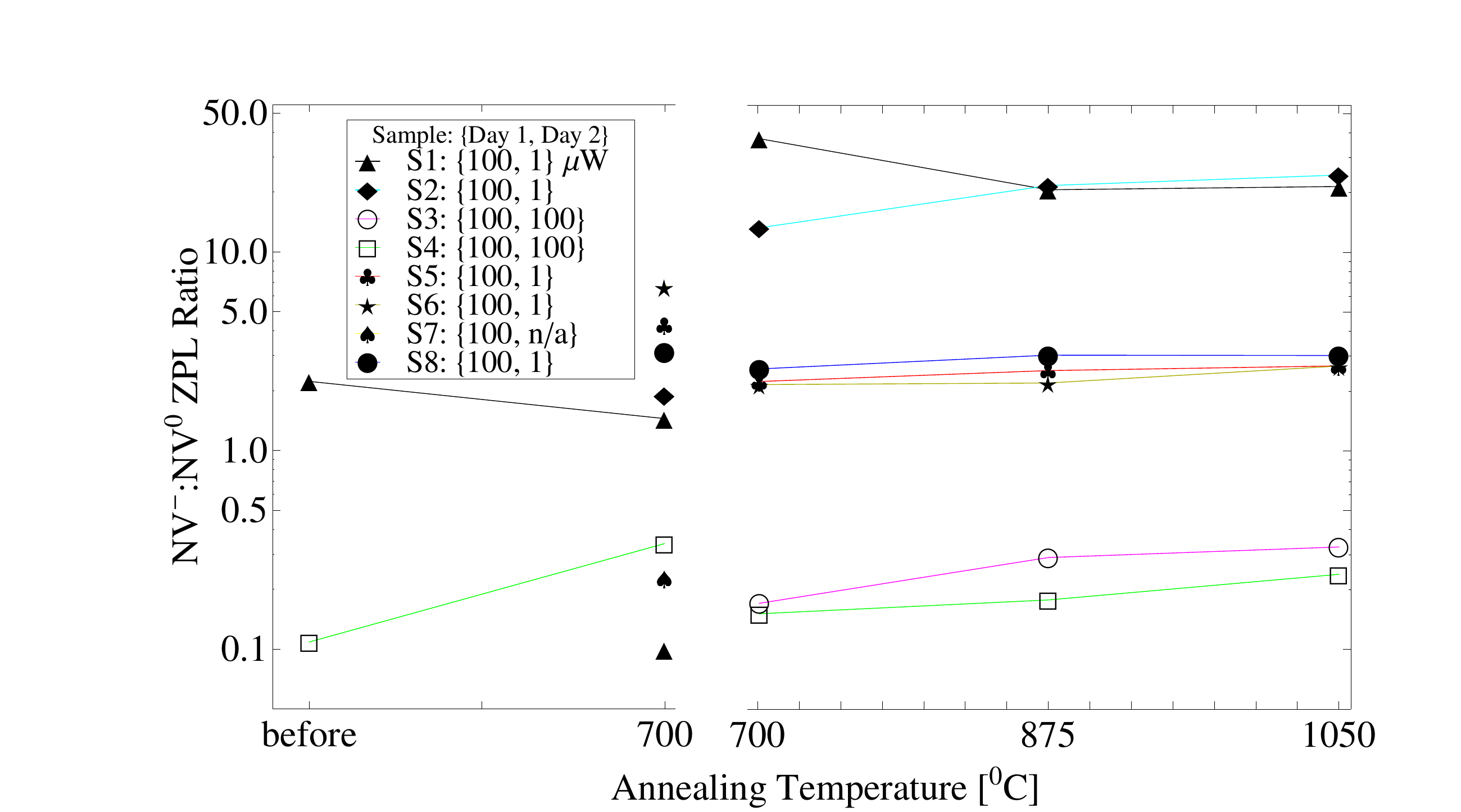}
        \vspace{1 cm}
        \begin{minipage}[b]{1\linewidth}
            \centering
            \textbf{(b)}
        \end{minipage}
        \vspace{-1 cm}
    \end{minipage}
    \caption{\label{fig:anneal}(Color online) (a) NV$^{\mbox{-}}$ concentrations of electron-irradiated samples as a function of annealing temperature. (b) NV$^{\mbox{-}}$:NV$^{0}$ ZPL intensity ratio of the same samples as a function of annealing temperature. The data were taken during two different runs on separate days. On each day the collection location for each sample was the same, but in between days the location changed. The laser power for the data on each day is indicated in the legend. Filled plot symbols are used for all samples with high nitrogen content and empty symbols represent the two samples with $\lesssim 1~{\rm ppm}$ nitrogen. Error bars are not shown here but are discussed in the text.}
\end{figure}

\begin{figure}
    \includegraphics[width = .88\textwidth]{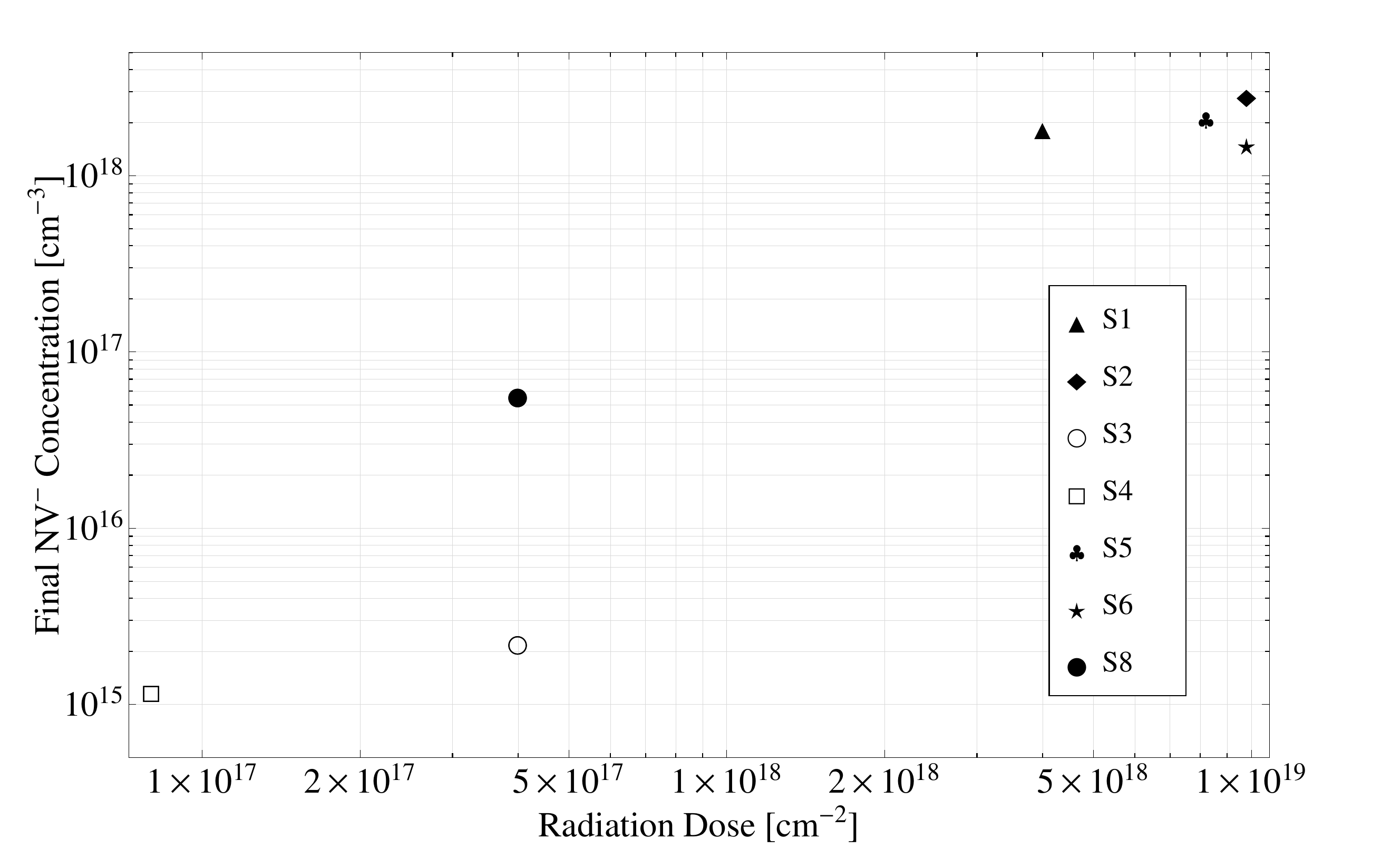}
    \caption{\label{fig:density}NV$^{\mbox{-}}$ concentrations near the surface as a function of dose for electron-irradiated samples after the final annealing at $1050^{\circ}$C. The data are the same as used in Fig. \ref{fig:anneal}(a). }
\end{figure}
Figure \ref{fig:anneal}(a) shows the NV$^{\mbox{-}}$ concentration for all of the electron-irradiated samples after each stage of annealing. All concentrations were determined by normalizing the integrated ZPL intensity by that of a single NV center, and the collection volume was $\sim 12~{\rm \mu m}$ from the surface. For the samples with high nitrogen content, the NV$^{\mbox{-}}$ concentration was already near, or at, its maximum value after the first annealing at $700~^{\circ}$C.  The samples with $\lesssim 1~{\rm ppm}$ nitrogen did not reach a maximum until the $875~^{\circ}$C annealing. We note that since we do not completely distinguish between the effects of annealing time and temperature, it is possible that annealing the low nitrogen-content samples at $700^{\circ}$C for twice as long would produce the same effect.

In order to test reproducibility, the data were taken on two different days, using different lateral locations in the sample, and in between days the mirrors and sample mount had been adjusted. The samples were analyzed after the $700^{\circ}$C (but before the $875^{\circ}$C) annealing on both days. The discontinuity seen in Fig. \ref{fig:anneal}(a) between days shows that the uncertainty in determination of the NV$^{\mbox{-}}$ concentrations is about a factor of two, caused by the aforementioned conditions. We estimate that the total systematic uncertainty due to radiation absorption, uncertainty in the collection volume, etc. is less than the spread in concentration due to spatial inhomogeneity.

The NV$^{\mbox{-}}$:NV$^{0}$ ZPL intensity ratio is plotted as a function of annealing temperature in Fig. \ref{fig:anneal}(b). The ZPL intensity ratio can be converted to a concentration ratio by using the difference in the Huang-Rhys factors (Sec. \ref{sec:photoionization}) for the two charge states (NV$^{\mbox{-}}$:NV$^{0}$ ZPL intensity ratio is approximately a factor of two smaller than the corresponding concentration ratio). The ratio does not appear to have a strong temperature dependence, but it varies strongly as a function of position in the sample and/or laser power, as evidenced by the large discontinuity between the two different days (corresponding to two different locations and laser powers). The incident laser powers are shown in the legend of Fig. \ref{fig:anneal}(b) because, as will be discussed in Sec. \ref{sec:photoionization}, changes in power lead to varying degrees of photo-ionization, which, depending on the individual sample characteristics, affects the ratio by an order of magnitude or more.

The very high NV$^{\mbox{-}}$:NV$^{0}$ ratio for Samples 1 and 2 provides evidence that many of the nitrogen atoms had not been converted to NV centers, since if this were the case, there would not be enough donors for making the negative charge state (Sec. \ref{sec:proton_discuss}). Conversely, the data in Fig. \ref{fig:anneal}(b) show the effects of electron depletion on the low nitrogen-content samples, leading to a significantly lower NV$^{\mbox{-}}$:NV$^{0}$ ratio. Since the specific collection location and laser power between days were different, there is a large degree of uncertainty in the absolute values of the NV$^{\mbox{-}}$:NV$^{0}$ ratio for each sample. However, for a given day and a given sample, the laser power and collection location were constant, so it can be concluded from the flat temperature dependence seen in Fig. \ref{fig:anneal}(b) for Day 2, and to a lesser extent also on Day 1, that the NV$^{\mbox{-}}$:NV$^{0}$ ratio is not strongly affected by annealing.

Figure \ref{fig:density} shows the NV$^{\mbox{-}}$ concentrations obtained after the final annealing at $1050^{\circ}$C as a function of radiation dose. As expected, the higher radiation dose generally leads to larger NV$^{\mbox{-}}$ concentrations. The initial nitrogen content is also an important factor, as evidenced by the factor of $\sim25$ difference between concentrations for Samples $3$ and 8. These samples were subjected to the same radiation dose but their initial nitrogen concentration differed by approximately a factor of $100$.

In conclusion, the doses and annealing temperatures listed in Tab. \ref{tab:prep} were sufficient to create samples with a wide range of NV$^{\mbox{-}}$ concentrations. Annealing at $700~^{\circ}$C was sufficient to obtain the maximum NV$^{\mbox{-}}$ yield for samples with high nitrogen content, while an additional annealing at $875~^{\circ}$C was required for the low-nitrogen-content samples. The high-nitrogen samples exhibited a large NV$^{\mbox{-}}$:NV$^{0}$ ratio, while the low-nitrogen samples had a significantly lower ratio. Lastly, the concentration-measurement technique based on photoluminescence was reproducible to within approximately a factor of two even when different lateral positions in the sample were used.

\subsubsection{Nitrogen-to-NV$^{\mbox{-}}$ conversion}
\label{sec:ntonv}
\begin{table}[ht]

\centering
    \begin{tabular}{c c c c c}
      \hline
      \hline
      \#~ & ~[N] (ppm)~ & ~post-irradiation [Vacancies] (ppm)~ & ~final [NV$^{\mbox{-}}$] (ppm)~ & ~[NV$^{\mbox{-}}$]:[N]~\\
      \hline
      1 & $\lesssim100$ & $45$ & $10$ & $\gtrsim10\%$ \\ 
      2 & $\lesssim100$ & $110$ & $16$ & $\gtrsim16\%$ \\ 
      3 & $\lesssim1$ & $5$ & $.012$ & $\gtrsim1\%$ \\ 
      4 & $\lesssim1$ & $1$ & $.007$ & $\gtrsim1\%$ \\ 
      5 & $\lesssim200$ & $90$ & $12$ & $\gtrsim6\%$ \\ 
      6 & $\lesssim200$ & $110$ & $8$ & $\gtrsim4\%$ \\ 
      8 & $\lesssim100$ & $5$ & $.3$ & $\gtrsim0.3\%$ \\ 
      [1ex]
      \hline
    \end{tabular}
    \caption{Sample Characteristics. Final NV$^{\mbox{-}}$ concentrations for the electron-irradiated samples are tabulated along with initial nitrogen content (reported by the commercial providers) and projected vacancy concentrations. These values are used to set a lower bound on the nitrogen-to-NV$^{\mbox{-}}$ conversion efficiency, as seen in the last column. The error in NV$^{\mbox{-}}$ concentration estimates, and consequently the error in conversion efficiency, is a approximately a factor of two, as discussed in the text.\label{tab:NtoNV}}
\end{table}

 A figure of merit for sensitive high-density magnetometry is the conversion efficiency of substitutional nitrogen to NV$^{\mbox{-}}$ centers, because any remaining paramagnetic nitrogen accelerates the rate of spin decoherence (see Sec. \ref{sec:T2}). For irradiation temperatures around those used in this work ($\lesssim 150^{\circ}$C), the rate of vacancy production for $2\mbox{-}{\rm MeV}$ electrons throughout millimeter-thick samples was determined experimentally to be $\sim{\rm 1.25~vacancies/e^{\mbox{-}}/cm}$ in Ref. \cite{NEW2002} and $\sim1.5 {\rm~vacancies/e^{\mbox{-}}/cm}$ in Ref. \cite{HUN2000} and theoretically to be $\sim{\rm 2.2~vacancies/e^{\mbox{-}}/cm}$ \cite{CAM2000}. Since the electrons are relativistic, the difference in vacancy production between 2- and $3\mbox{-}{\rm MeV}$ electrons is small \cite{CAM2000}, so we assume a vacancy production for the $3\mbox{-}{\rm MeV}$ electrons used in this work of $\sim2~{\rm vacancies/e^{\mbox{-}}/cm}$.  Using this value, along with irradiation doses from Tab. \ref{tab:prep}, Tab. \ref{tab:NtoNV} displays approximate initial concentrations of vacancies before annealing. The initial nitrogen concentrations reported by the manufacturers and the final NV$^{\mbox{-}}$ concentrations from Fig. \ref{fig:density} are also shown and these values are used to set a lower bound on the nitrogen-to-NV$^{\mbox{-}}$ conversion efficiency. Sample 2, having converted $\gtrsim16\%$ of the initial nitrogen into NV$^{\mbox{-}}$ centers, shows the most promising conditions for high conversion efficiency. For this sample, the initial nitrogen concentration was $\lesssim 100~{\rm ppm}$ and the irradiation dose was $9.8\times10^{18}~{\rm cm^{-2}}$, corresponding to $\sim110~{\rm ppm}$ vacancies. As discussed in Sec. \ref{sec:absorption}, it is likely this dose was insufficient to maximize NV$^{\mbox{-}}$ formation.

\subsubsection{Visible absorption}
\label{sec:absorption}
Optical magnetometry with spin ensembles requires that the samples be reasonably transparent over the spectral range of interest. In diamond magnetometry based on NV$^{\mbox{-}}$ centers, zero-phonon optical transitions at $638~{\rm nm}$ and $1046~{\rm nm}$ \cite{ROG2008}, as well as their broad phonon sidebands, provide possible channels for detection. To measure the opacity in this range, a Cary 50 spectrophotometer was used to obtain transmission spectra for four of the electron-irradiated samples (Samples $1-4$) after the final annealing at $1050^{\circ}$C. The visible spectra for a subset of these samples were also measured after the first annealing, and no significant differences in the spectra between these stages of annealing were observed.

\begin{figure}
    \includegraphics[width = .88\textwidth]{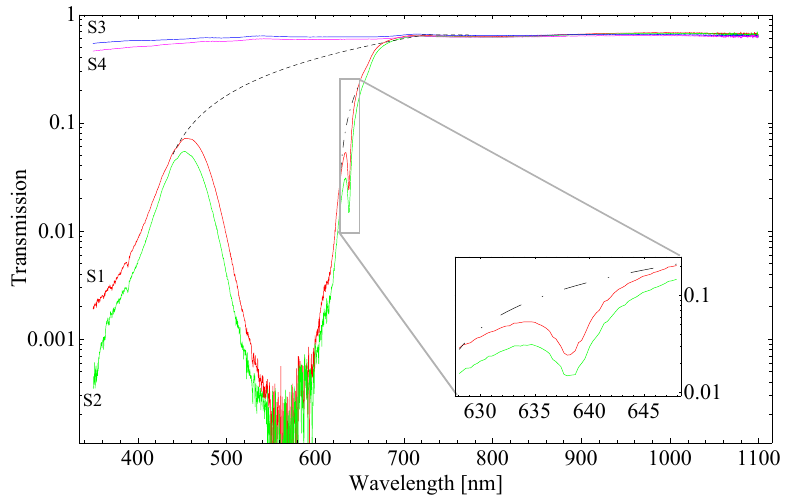}
    \caption{\label{fig:absorption}(Color online) Room temperature transmission curves for four electron-irradiated samples after the final annealing at $1050^{\circ}$C. The beam path (i.e. the sample thickness) was $1~{\rm mm}$ for Samples $1$ and 2 (S1 and S2) and $.5~{\rm mm}$ for S3 and S4. The ZPL at $\sim 638~{\rm nm}$ is visible for S1 and S2 and is blown up in the inset. The dotted lines represent the baselines used for Sample $1$ in the effective vibrational-splitting and Huang-Rhys factor calculations (the baseline for Sample $2$ was similar).}
\end{figure}

Figure \ref{fig:absorption} shows the room-temperature transmission curves for these samples. For all four samples, the transmission in the wavelength region $\sim700\mbox{-}1100~{\rm nm}$ is approximately $66\%$. The\ loss of light can be accounted for by Fresnel reflection; the index of refraction of diamond is 2.4, which corresponds to a reflection of $17\%$ off of each diamond surface for light at normal incidence in air.

In the wavelength region to the red of the ZPL and below $\sim700~{\rm nm}$, Samples $1$ and 2 exhibit absorption that increases rapidly with decreasing wavelength.  Absorption at these wavelengths is due to the significant population of NV$^{\mbox{-}}$ centers in the excited vibrational states at room temperature. After subtraction of the overall background and absorption due to the ZPL (baselines in Fig. \ref{fig:absorption}), the ratio of the integrated absorption in this spectral region ($638\mbox{-}700~{\rm nm}$) to the total integrated absorption is $\sim11(2)\%$ for Sample $1$ and $\sim13(2)\%$ for Sample $2$. If these values are taken to be the Boltzmann-distributed occupancy of excited vibrational states at room temperature, they would give a rough estimate for the effective ground-state vibrational-energy splitting \cite{SAN2009NOTE} of $56(4)$ and $52(4)~{\rm meV}$ for the NV$-$ center, respectively. These values are in tolerable agreement with previous measurements based on emission spectra, where the effective vibrational-spacing in the $^3A_2$ ground state of the NV$^{\mbox{-}}$ center was estimated to be $\sim60~{\rm meV}$ \cite{DAV1976,KIL2000}.

In the region around $\sim628-648~{\rm nm}$, inset in Fig. \ref{fig:absorption}, the NV$^{\mbox{-}}$ ZPL appears prominently in the spectrum for Samples $1$ and 2, where at the peak of absorption approximately $98\%$ of the light is absorbed. This corresponds to four absorption lengths for this $1\mbox{-}{\rm mm}$ path length, with approximately one absorption length attributed to just the ZPL and the other three due to overlap of the PSB. To the blue of the ZPL, at $\sim450 - 638~{\rm nm}$, resides a broad feature which is a combination of the NV$^0$ and NV$^{\mbox{-}}$ phonon sidebands. The ZPL for NV$^0$ at $575~{\rm nm}$ is obscured by this strongly absorbing region. For Samples $3$ and 4, there are no visible spectral features, due to the low NV concentrations in these samples (see Tab. \ref{tab:NtoNV}).

Based on the absorption data, it is possible to estimate the electron-phonon coupling parameter, also known as the Huang-Rhys factor \cite{DAV1974}, $S=-\ln(I_{ZPL}/I_{tot})$, where $I_{ZPL}$ and $I_{tot}$ refer to the integrated absorption intensities after normalization by the light frequency. While this relationship is only strictly correct at zero-temperature where quadratic electron-phonon coupling is suppressed, it was shown in Ref. \cite{DAV1974} that, in NV$^{\mbox{-}}$ centers, the temperature-dependence of $S$ is weak. In that work it was reported that $S$ increases by less than $30\%$ from 2 to $350~{\rm K}$. Analysis of the data in Fig. \ref{fig:absorption} gives, after subtracting the linear background (dashed lines in the figure), $S\approx3.9$ for Sample $1$ and $4.0$ for Sample $2$, which is within the range of values in the literature for NV$^{\mbox{-}}$ centers at or below $80~{\rm K}$ ($4.8, 4.3, 2.8$ in Ref. \cite{KIL2000}, and $3.7$ in Ref. \cite{DAV1976}). The value used for $I_{tot}$ is a slight over-estimate, as the phonon sideband from the NV$^0$ centers overlaps, but this effect is small ($\lesssim2\%$) since the concentration of NV$^0$ was small compared to that of NV$^{\mbox{-}}$ for both samples (see Sec. \ref{sec:NVconcentrations}).

Another important observation from Fig. \ref{fig:absorption} is the absence of an absorption feature at $741~{\rm nm}$, the wavelength of the ZPL of the neutral vacancy center, GR1. The neutral vacancy concentration has previously been calibrated based on the GR1 integrated ZPL in absorption \cite{LAW1998,TWI1999}. Based on this calibration and the lack of an identifiable ZPL, a conservative upper bound of $1~{\rm ppm}$ can be placed on the concentration of neutral vacancy centers for Samples 1 and 2. This suggests that the NV conversion efficiency for these samples may be limited by the number of vacancies present after irradiation. Recall from Tab. \ref{tab:NtoNV} that the vacancy concentration was predicted to be slightly larger than the nitrogen concentration for Sample $2$. Since all of the neutral vacancies are gone after the final annealing, it is likely that there is a competing process to the vacancy capture by the substitutional nitrogen atoms.

We also considered the possibility that many vacancies remain after annealing, but the vast majority of remaining vacancies were actually negatively charged, due to the high concentration of nitrogen donors, and therefore do not contribute to the GR1 absorption. The negatively charged vacancy center, ND1, is not visible in the absorption spectra because its ZPL is at $396~{\rm nm}$ \cite{ZAI2001}. This wavelength coincides with the large background absorption, presumably due to photo-ionization of the substitutional nitrogen, prohibiting a direct measurement of the negatively charged vacancy concentration. However, as will be shown in Sec. \ref{sec:IRabsorption}, the concentration of positively charged nitrogen, the likely product of ND1 formation, can be measured from infrared absorption spectra and was found to be $15(3)$ and $19(4)~{\rm ppm}$ after the final annealing for Samples 1 and 2, respectively. Since much of this concentration can be attributed to nitrogen's role as a donor in NV$^{\mbox{-}}$ formation (recall that the NV$^{\mbox{-}}$ concentrations were approximately 10 and $16~{\rm ppm}$ for Samples 1 and 2, respectively), a conservative upper bound of $10~{\rm ppm}$ can be placed on the negatively charged vacancy concentration in both samples. Since this upper bound represents a small fraction of the estimated initial vacancy concentration (\ref{tab:NtoNV}), it remains likely that in order to maximize the NV conversion efficiency higher radiation doses should be explored.

For Samples 1 and 2, the visible-absorption measurements can also be used as a cross-check on the NV$^{\mbox{-}}$ concentrations determined from the PL spectra (Sec. \ref{sec:NVconcentrations}) \cite{DAV1992,LAW1998}. Using the calibration published in Ref. \cite{LAW1998}, the integrated ZPLs for Samples 1 and 2 correspond to NV$^{\mbox{-}}$ concentrations of 8 and $10~{\rm ppm}$, respectively, which is in tolerable agreement with the concentrations measured from the PL spectra (10 and $16~{\rm ppm}$, respectively). As the calibration established in that work was intended for low-temperature absorption, the somewhat lower [NV$^{\mbox{-}}$] values can be partly attributed to the slightly lower Huang-Rhys factor at room-temperature \cite{DAV1974}.

In summary, visible-absorption measurements establish that even at high radiation doses the dominant source of absorption in the visible and near-infrared is due to NV centers, and not other radiation damage. The absence of GR1 absorption suggests that higher radiation doses should be explored for optimizing NV yield. The absorption spectra were also used to confirm the NV$^{\mbox{-}}$ ground vibrational-state occupancy and Huang-Rhys factors from the literature and the NV$^{\mbox{-}}$ concentrations determined from the absorption ZPL provide good agreement with those determined from the PL spectra in Sec. \ref{sec:NVconcentrations}.

\subsubsection{Infrared absorption}
\label{sec:IRabsorption}

\begin{figure}
\centering
    \includegraphics[width = .65\textwidth]{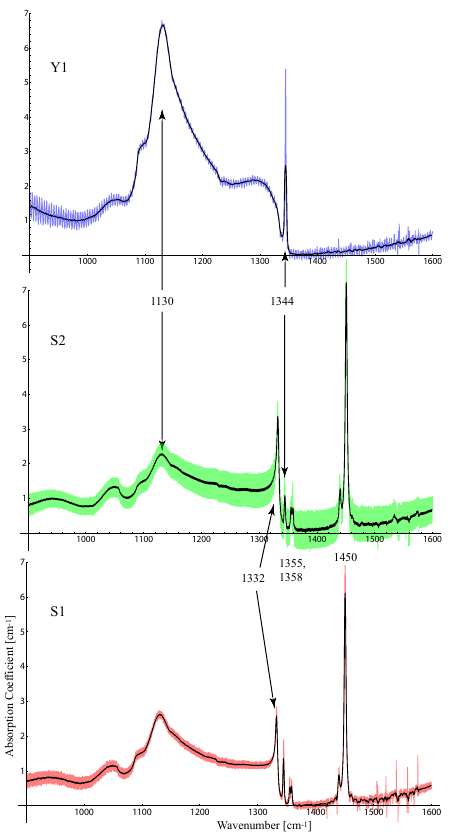}
    \caption{\label{fig:ftir}(Color online) Infrared spectra for Samples $1$, 2, and ``Y1'' after the final annealing. Each spectrum represents averaging of 64 scans and the resolution of the spectrometer was set to $0.25~{\rm cm}^{-1}$. The rapid oscillations are due to etaloning at normal incidence, and they have been smoothed by a moving average (black lines) in order to quantify the peak at $\sim1130~{\rm cm}^{-1}$ (see text). Features mentioned in the text are labeled by their wavenumbers.}
\end{figure}

The presence of paramagnetic nitrogen has important implications for magnetometry, as discussed below in Sec. \ref{sec:T2}. Infrared absorption measurements were made with a Nicolet Magna IR-750 FTIR spectrometer to measure the leftover nitrogen concentration after the final annealing (Fig. \ref{fig:ftir}). Nitrogen-rich diamonds display a broad vibrational absorption band peaking around $1130~{\rm cm}^{-1}$ \cite{COL1982} that has been calibrated with respect to the substitutional nitrogen concentration using ESR and inert-gas fusion techniques \cite{CHR1971,WOO1990}. According to this calibration, an absorption coefficient of $1~{\rm cm}^{-1}$ corresponds to a substitutional nitrogen concentration of approximately $22~{\rm ppm}$ \cite{WOO1990}. Figure \ref{fig:ftir} shows the infrared spectra for Samples 1 and 2 after the final annelaing, and a third Sumitomo HPHT diamond (``Y1'') with very similar specifications that had not been irradiated or annealed. Based on the spectral feature at $\sim1130~{\rm cm}^{-1}$, the nitrogen content for Samples 1, 2, and Y1 were $58(8),~49(7),~{\rm and}~146(20)~{\rm ppm}$, respectively, where the uncertainty is based on the spread of the calibration values in the literature \cite{CHR1971,WOO1990}. The sharp spectral feature at $1344~{\rm cm}^{-1}$ present in all three samples is also associated with the single substitutional nitrogen \cite{COL1982}. Both of these peaks are more intense for Sample Y1, presumably because Samples 1 and 2 contain NV centers and other nitrogen-related centers formed during irradiation and annealing. The most prominent feature in the FTIR spectra for Samples 1 and 2 is at $1450~{\rm cm}^{-1}$, which was tentatively assigned to a vibrational mode of interstitial nitrogen \cite{WOO1982,KIF1996}. The smaller features at $1355$ and $1358~{\rm cm}^{-1}$ may also be related to this defect \cite{COL1987}. The production of this interstitial defect may be due to irradiation, since these features are not present in Sample Y1.

The spectral feature at $1332~{\rm cm}^{-1}$, present only in Samples 1 and 2, has been linked to positively charged substitutional nitrogen, N$^+$ (Ref. \cite{LAW1998}). In that work, it was argued that because of nitrogen's role as a donor in NV$^{\mbox{-}}$ formation, the concentration of N$^+$ after the final annealing should be roughly equal to the concentration of NV$^{\mbox{-}}$. Using the calibration of the $1332~{\rm cm}^{-1}$ peak from that work ($1~{\rm cm}^{-1}$ of absorption corresponds to $5.5\pm1~{\rm ppm}$ of N$^+$), the NV$^{\mbox{-}}$ concentrations for Samples 1 and 2 would be $15(3)$ and $19(4)~{\rm ppm}$. Since nitrogen is a donor in the formation of other centers, such as the negatively charged vacancy (Sec. \ref{sec:absorption}), it can be expected that [N$^+$] would be somewhat higher than [NV$^{\mbox{-}}$]. Thus these values for [NV$^{\mbox{-}}$] are in excellent agreement with the values ($10$ and $16~{\rm ppm}$, respectively) measured using PL spectra in this work.

In conclusion, the FTIR measurements indicate that a large concentration of paramagnetic nitrogen remains in the samples even after the highest irradiation doses and annealing temperatures studied here. The absorption of the $1332~{\rm cm}^{-1}$ peak was found to be in excellent agreement with the NV$^{\mbox{-}}$ concentrations determined from PL spectra.

\subsection{Light-power dependence of NV$^0$ and NV$^{\mbox{-}}$ concentrations}
\label{sec:photoionization}
\begin{figure}[ht]
    \begin{minipage}[b]{1\linewidth}
        \centering
        \includegraphics[width=.65\textwidth]{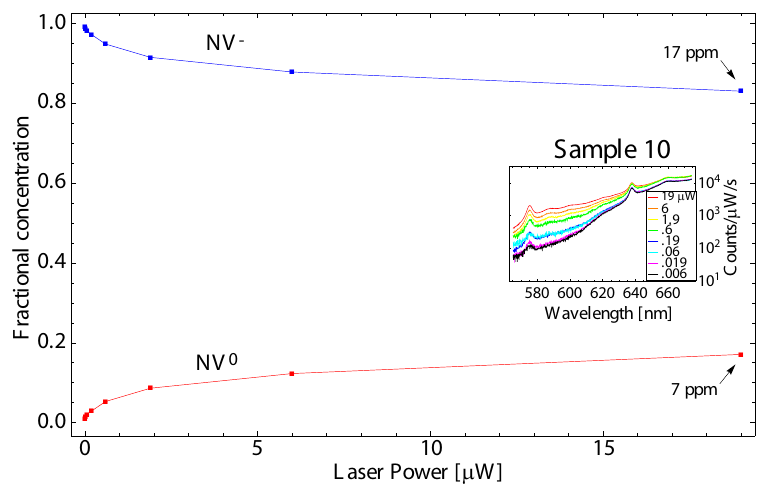}
        \vspace{1 cm}
        \begin{minipage}[b]{1\linewidth}
            \centering
            \textbf{(a)}
        \end{minipage}
        \vspace{-1.5 cm}
    \end{minipage}
    \begin{minipage}[b]{1\linewidth}
        \centering
        \includegraphics[width=.65\textwidth]{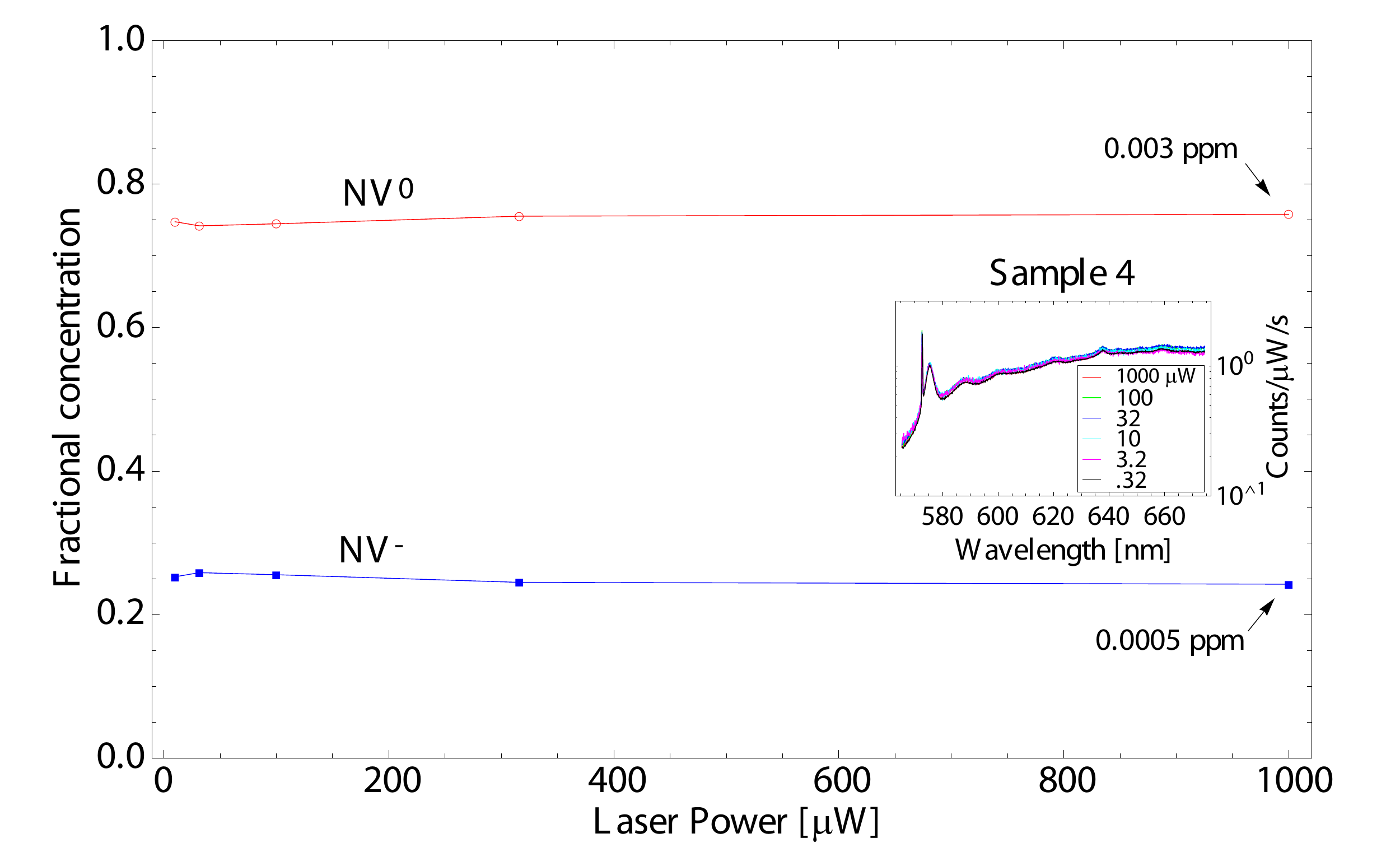}
        \vspace{1 cm}
        \begin{minipage}[b]{1\linewidth}
            \centering
            \textbf{(b)}
        \end{minipage}
        \vspace{-1 cm}
    \end{minipage}
    \caption{\label{fig:photoionization}(Color online) Relative NV$^{\mbox{-}}$ and NV$^0$ concentrations for (a) Sample $10$ and (b) Sample 4 after the first annealing ($700^{\circ}$C and $800^{\circ}$C, respectively) as a function of incident laser power. The concentrations were obtained from fluorescence spectra (insets) taken at a depth of approximately $36~{\rm \mu m}$, for Sample $10$, and $12~{\rm \mu m}$, for Sample $4$. Determination of NV$^0$ concentration was made by weighting the relative ZPL intensity according to the difference in room-temperature Huang-Rhys factors; $S\sim3.3$ for NV$^0$ \cite{ZAI2001} and $S\sim4.0$ for the NV$^-$ center (Sec. \ref{sec:absorption}). The laser beam was focused to a diameter of approximately $0.5~{\rm \mu m}$ for all measurements.}
\end{figure}
The neutral NV center, NV$^{0}$ presents a similar problem to ultra-sensitive magnetometry as the remaining substitutional nitrogen, since it is also paramagnetic \cite{FEL2008} and therefore limits NV$^{\mbox{-}}$ dephasing times. Further, since the emission spectrum of NV$^0$ partly overlaps with the phonon sideband of the NV$^-$ defect, the presence of NV$^0$ leads to a background fluorescence signal that reduces measurement contrast. In previous work \cite{MAN2005,GAE2006APB}, it was shown that the NV$^{\mbox{-}}$ and NV$^0$ concentrations can change upon illumination by light in a range of wavelengths including $532~{\rm nm}$, the wavelength used for excitation in this work. In particular, in Ref. \cite{MAN2005} it was reported that the NV$^{\mbox{-}}$:NV$^0$ ratio dropped with increasing laser power and identified direct photo-ionization of the NV$^{\mbox{-}}$ center and ionization of nearby potential nitrogen donors as the primary causes.

Figure \ref{fig:photoionization} shows the relative NV$^{\mbox{-}}$ and NV$^0$ concentrations determined from the PL spectra (Sec. \ref{sec:fluoresce}) for (a) Sample $10$ and (b) Sample 4 as a function of incident laser power. By relative, we mean that the NV$^{\mbox{-}}$ and NV$^0$ concentrations have been normalized such that, at any given laser power, their sum is 1. In order to determine the NV$^0$ concentration, the difference in Huang-Rhys factors for NV$^0$ and NV$^{\mbox{-}}$ were used for normalization, since the NV$^0$ ZPL for the single NV center (Fig. \ref{fig:fluorescence}(a)) was not visible in the spectrum. As seen in Fig. \ref{fig:photoionization}, the laser-light intensity has a significant effect on the NV charge state for the high-nitrogen-concentration Sample $10$, but does not affect the low-nitrogen-concentration Sample $4$. Since at very low light power Sample 10 has a large [NV${\mbox{-}}$]:[NV$^0$] ratio and Sample 4 has a very low ratio, the results in Fig. \ref{fig:photoionization} can be interpreted as a small fraction of NV$^{\mbox{-}}$ being converted to NV$^0$ due to the laser light until the ratio reaches some minimum saturation level. Note that the range of laser powers explored for Sample 4 was higher than that of Sample 10, due to the low overall NV concentration in this sample. Similar plots were made after each annealing, but further annealing was found to have no measurable impact. While the mechanism for changing the NV charge state (i.e. one- or two- photon process) cannot be confirmed from this data set, the results indicate that sensitive magnetometry requires careful optimization of the laser light power.

\subsection{Optically detected magnetic resonance (ODMR)}
\label{sec:odmr}
\subsubsection{NV$^{\mbox{-}}$ electronic structure}
\label{sec:energystructure}
\begin{figure}
\centering
    \includegraphics[width=.85\textwidth]{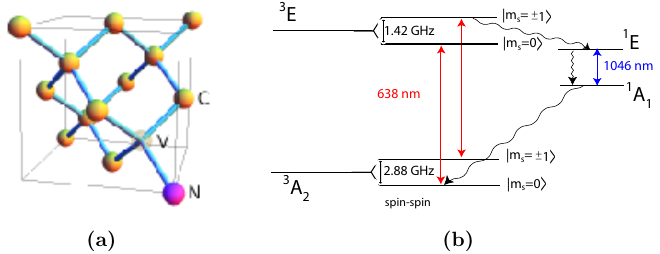}
    \caption{\label{fig:structure}(Color online) (a) Crystallographic cell for the NV center in diamond. The center has C$_{3v}$ symmetry and the line that contains both N and V defines the symmetry axis, of which there are four possible orientations. C--Carbon, N--Nitrogen, V--Vacancy. (b) Level diagram for NV$^{\mbox{-}}$ center showing spin-triplet ground and excited states, as well as the singlet system responsible for intersystem crossing (see text). The transition wavelengths are for the ZPL, with radiative transitions indicated by solid arrows and non-radiative transitions by wavy arrows.}
\end{figure}

One of the crucial features of nitrogen-vacancy ensembles for practical magnetometry is that the NV$^{\mbox{-}}$ centers can be optically pumped and interrogated using visible light over a broad range of wavelengths (see Fig. \ref{fig:absorption}). Figure \ref{fig:structure}(a) shows the crystallographic cell for the trigonal NV$^{\mbox{-}}$ center in one of four possible orientations, corresponding to the N-V axis along each of the dangling bonds. Figure \ref{fig:structure}(b) shows the level diagram with energy levels and allowed transitions labeled according to the most current model posited in Refs. \cite{ROG2008,ROG2009}. Based on C$_{3v}$ symmetry \cite{DAV1976} and results from early electron-spin-resonance (ESR) \cite{LOU1978} and hole-burning experiments \cite{RED1987}, it is known that the center's ground state is a $^3A_2$ spin-triplet and the state which can be directly accessed by an electric dipole transition is $^3E$. The ground state is split by $2.88~{\rm GHz}$ due to the strong spin-spin interaction arising from the electron charge distribution being highly localized near each of the three carbon atoms \cite{LOU1978}. At low temperature the $^3E$ state is an orbital doublet split by transverse strain, but at room temperature the orbital components appear to average \cite{ROG2009}, and thus it is represented by a single orbital with spin-spin splitting of $1.42~{\rm GHz}$ \cite{BAT2009}. The $^3A_2$ ground state has a single non-degenerate electronic orbital and thus is immune to first-order shifts due to strain.

In thermal equilibrium, the ground state sublevels are nearly equally populated, but after interaction with light resonant with the $^3A_2 \rightarrow{^3E}$ transition (excited here on the phonon sideband at $532~{\rm nm}$), the $|m_s = \pm 1\rangle$ sublevels become depopulated and the $|m_s = 0\rangle$ sublevel eventually takes on the excess. Theoretical considerations and spin-dependent lifetime measurements suggest that the process responsible for the ground-state polarization is the decay, due to spin-orbit interaction, from the $|m_s = \pm 1\rangle$ sublevels of the $^3E$ to a singlet state(s) with lower energy \cite{JEL2006,MAN2006,BAT2008}. Pulsed-pump transient measurements have shown that the $|m_s=\pm1\rangle$ can decay back to the $|m_s=0\rangle$ through the singlet system after approximately $300~{\rm ns}$ \cite{MAN2006}. As this decay is mostly non-radiative, optical pumping into the $|m_s=0\rangle$ sublevel results in an increase in fluorescence with a contrast as high as $30\%$, limited by the branching ratio of approximately 0.5 for the $|m_s=\pm1\rangle$ sublevels decay to the singlet system \cite{BAL2008}. Until very recently this singlet system, thought to be $^1A$ and/or $^1E$ from $C_{3v}$ group theory considerations \cite{MAN2006,ROG2008}, had not been directly detected. Weak infrared emission at $1046$ nm was finally observed in Ref. \cite{ROG2008} and the magnetic-field and strain dependence of the fluorescence spectra indicate that the emission is indeed involved in the optical pumping mechanism.

\subsubsection{Optically detected magnetic resonance technique}
\begin{figure}
\centering
    \includegraphics[width=.8\textwidth]{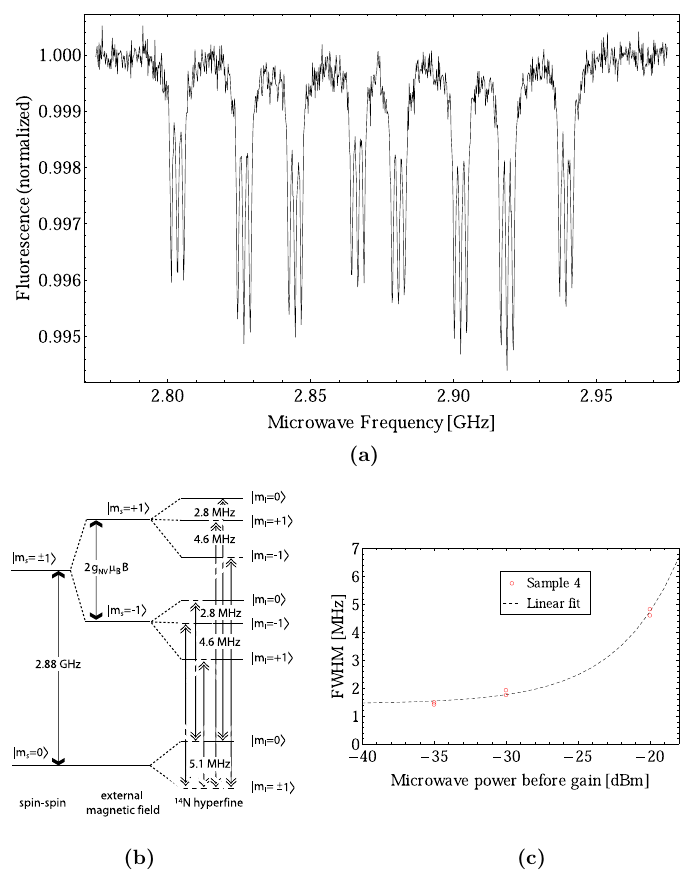}
    \caption{\label{fig:odmr}(Color online) (a) ODMR spectrum for Sample $3$. The 24 separate resonances correspond to two different $|\Delta m_s|=1$ transitions, four different NV orientations, and three different $^{14}$N hyperfine transitions. The orientations were split by a $\sim50$ G field. (b) Level diagram for the ground-state magnetic dipole transitions induced by the microwave field in the ODMR measurements. The selection rules for these transitions are $|\Delta m_s|=1$ and $\Delta m_I=0$. The values for the $^{14}$N hyperfine splitting are taken from Ref. \cite{RAB2006}. (c) Microwave-power dependence (log-scale) of the relaxation rate for Sample $4$ for a laser power of $213~{\rm \mu W}$. Microwaves from a signal generator were amplified and applied with a $\sim1\mbox{-}{\rm mm}$-diameter current loop with unmatched impedance. The microwave power was measured at the output of the signal generator prior to the $\sim30~{\rm dB}$ amplifier. The FWHM linewidth of each $|\Delta m_s|=1, \Delta m_I=0$ transition for one of the orientations was plotted as a function of microwave power. A linear fit for the lowest-power points (dashed line) was used to extrapolate the linewidth to zero microwave power. The standard error in the fit parameter serves as the uncertainty in the measurement.}
\end{figure}

The transverse spin relaxation time, $T_2^{\ast}$, is critically important for spin-precession magnetometry, as it places bounds on both the minimum sensitivity and laser power of broadband sensors (Sec. \ref{sec:prospects}). In order to measure $T_2^{\ast}$, the technique of ODMR was employed after the final annealing. In this technique, a microwave field is tuned to the ground-state ESR frequency of the NV$^{\mbox{-}}$ center and the fluorescence resulting from illumination by light exciting the $^3A_2 \rightarrow{^3E}$ transition is monitored \cite{LOU1978,HE1993A}. Optical pumping transfers the NV$^{\mbox{-}}$ centers to the $|m_s=0\rangle$ ground state. An external magnetic field is applied, and the electronic spin sublevels $|m_s = \pm 1\rangle$ couple to the component $B_z$ along the N-V axis, leading to Zeeman energy shifts of $\Delta E = m_s g_{s}\mu_B B_z$, where $g_{s}=2.0028$ is the NV$^{\mbox{-}}$ electronic Land\'e factor \cite{LOU1978} and $\mu_B\approx1.40~{\rm MHz/G}$ is the Bohr magneton. When the oscillating magnetic field has frequency equal to the splitting between the $|m_s=0\rangle$ and $|m_s = \pm 1\rangle$ magnetic sublevels, the component transverse to the N-V axis drives the magnetic dipole transition and repopulates the $|m_s = \pm 1\rangle$ magnetic sublevels, counteracting the anisotropy produced by optical pumping. With the pump light still on, the increase in population in $|m_s=\pm1\rangle$ leads to non-radiative decay through the singlet system, reducing the fluorescence emitted by the center. Thus the ESR signal is the change in fluorescence registered when scanning the frequency of the microwave field.

An example of an ODMR spectrum is shown in Fig. \ref{fig:odmr}(a). An additional layer of complexity in the spectrum is the presence of the hyperfine interaction with the nuclear-spin-one $^{14}$N nucleus, which splits each $|\Delta m_s|=1$ resonance into three hyperfine peaks ($\Delta m_I=0$) with an equal spacing of $2.3~{\rm MHz}$ at zero-magnetic field \cite{HE1993A}. Figure \ref{fig:odmr}(b) shows the level diagram for the allowed ground-state transitions. The four different orientations of the NV$^{\mbox{-}}$ center are degenerate in the absence of magnetic field, but this degeneracy was removed with a static magnetic field of magnitude $\approx50~{\rm G}$, applied using a permanent magnet. Both fits to the data and numerical diagonalization of the $9\times9$ spin hamiltonian \cite{RAB2006} confirmed that the spin-state mixing due to this applied field, particularly the component perpendicular to the N-V axes, causes the hyperfine splitting to vary by $\lesssim 0.1~{\rm MHz}$.

\subsubsection{Results of ODMR measurements}
\begin{table}[ht]
\centering
    \begin{tabular}{c c c c c c}
      \hline
      \hline
      Sample \#~ & ~[N] (ppm)~ & ~final [NV$^{\mbox{-}}$] (ppm)~ & ~Laser Power (${\rm \mu W}$)~ & ~$T_2^{\ast}$ (ns)~\\
      \hline
      1 & $\lesssim100$ & $10$ & $5$ & $66(8)$ \\ 
      2 & $\lesssim100$ & $16$ & $188$ & $118(48)$ \\ 
      3 & $\lesssim1$ & $.012$ & $130$ & $\gtrsim291$ \\ 
      4 & $\lesssim1$ & $.007$ & $213$ & $222(11)$ \\ 
      5 (spot 1) & $\lesssim200$ & $1.0$ & $24$ & $27(4)$ \\ 
      5 (spot 2) & $\lesssim200$ & $.02$ & $100$ & $128(64)$ \\ 
      5 (spot 3) & $\lesssim200$ & $2.5$ & $240$ & $145(63)$ \\ 
      8 & $\lesssim100$ & $.3$ & $188$ & $114(25)$ \\ 
      [1ex]
      \hline
    \end{tabular}
    \caption{Effective transverse spin relaxation times, $T_2^{\ast}$, extrapolated to zero microwave power, for several electron-irradiated samples at the given laser powers (focused to $\sim0.5\mbox{-}{\rm \mu m}$ diameter). For Sample $5$, three locations, corresponding to spots in different growth-sectors, were investigated. The uncertainty estimates for all samples except Sample $3$ come from the standard error of the linear fits, as shown in Fig. \ref{fig:odmr}(c). For Sample $3$, only one low microwave power, $-30~{\rm dBm}$ before the amplifier, was investigated (shown in Fig. \ref{fig:odmr}(a)). The lower bound on $T_2^{\ast}$ for this sample corresponds to one standard deviation below the statistical average of the resonance widths. \label{tab:T2}}
\end{table}

The resonance profile corresponding to each N-V orientation and $|\Delta m_s|=1$ coherence was fit with three Lorentzian functions of equal amplitude and equal linewidth, $\gamma$, given as full-width at half maximum (FWHM). Equal spacings between the centers of adjacent Lorentzians of $2.2~{\rm MHz}$ were used, even when the $^{14}$N hyperfine splitting could not be resolved. The effective transverse spin relaxation times, $T_2^{\ast}=1/(\pi \gamma)$, and associated uncertainties were obtained from the ODMR spectra by extrapolating $\gamma$ to zero microwave power (Fig. \ref{fig:odmr}(c)) with a linear fit. Density-matrix calculations confirmed that $\gamma$ is, to lowest order, linear in the microwave power. The resulting $T_2^{\ast}$ times are tabulated in Tab. \ref{tab:T2} for each sample at the given laser power. The laser-power dependence was not studied here though the light power was varied from sample to sample in order to maximize the signal on the photon counter. This dependence will be investigated in future work.

\subsubsection{Contributions to $T_2^{\ast}$}
\label{sec:T2}
For the HPHT-synthesized Sample $5$, ODMR spectra at three different locations, corresponding to sectors with very different initial nitrogen concentrations, were observed, and the NV$^{\mbox{-}}$ concentrations at these locations were estimated based on the integrated ZPL, as described in Secs. \ref{sec:fluoresce} and \ref{sec:NVconcentrations}. Since the irradiation was uniform throughout the sample, the wide variation of $T_2^{\ast}$, from $27~{\rm ns~to}~145~{\rm ns}$ is probably not due to vacancy-related defects other than NV centers, unless the formation of such centers is strongly correlated with the local concentration of nitrogen.

For dilute spins, the contribution of NV$^{\mbox{-}}$--NV$^{\mbox{-}}$ dipolar interactions to the magnetic-resonance broadening can be approximated by assuming that each NV$^{\mbox{-}}$ center couples to only the nearest-neighboring NV$^{\mbox{-}}$ center. For an ensemble, this dipolar coupling leads to a spin-relaxation contribution on the order of $T_{NV}\approx 1/[(g_s \mu_B)^2 n_{NV}]$, where $n_{NV}$ is the NV$^{\mbox{-}}$ concentration \cite{TAY2008}. For Sample 2, this corresponds to $T_{NV}\approx1~\mu{\rm s}$, meaning that NV$^{\mbox{-}}$--NV$^{\mbox{-}}$ coupling is significant but not necessarily the primary source of relaxation. For Sample 5, spot 3 has a much higher NV$^{\mbox{-}}$ concentration but similar $T_2^{\ast}$ to spot 2, providing evidence that NV$^{\mbox{-}}$--NV$^{\mbox{-}}$ relaxation is not the dominant source of relaxation in this sample.

For nitrogen-to-NV$^{\mbox{-}}$ conversion efficiency of less than $50\%$, the contribution of substitutional-nitrogen dipolar interactions to NV$^{\mbox{-}}$ decoherence is expected to be even larger, as the paramagnetic nitrogen has similar coupling strength. Similarly to the estimate for NV$^{\mbox{-}}$--NV$^{\mbox{-}}$ interactions, the characteristic timescale for this decoherence is $T_{N}\approx1/((g_s \mu_B)^2 n_{N})$, where we ignore the small difference in g-factors between NV$^{\mbox{-}}$ and nitrogen \cite{LOU1978}. Recall from Sec. \ref{sec:IRabsorption} that Samples 1 and 2 had final nitrogen concentrations on the order of $50~{\rm ppm}$, giving a value of $T_N\approx300~{\rm ns}$. Samples 1, 2, and 5-8 all had similar initial concentration of nitrogen ($\lesssim100~{\rm ppm}$). Since the irradiation and annealing process converted no more than $16~{\rm ppm}$ of nitrogen into NV$^{\mbox{-}}$ centers, it is likely that relaxation due to remaining paramagnetic nitrogen is the dominant cause of relaxation in all of these samples.

Another well-known cause for relaxation is inhomogenous broadening due to dipolar hyperfine coupling of nearby $^{13}$C nuclear spins (natural $1.1\%$ abundance in all samples). Based on experiments on single NV centers, this spin-relaxation contribution is in the range $T_{C_{exp}}\approx 0.3 - 1.5~\mu{\rm s}$ \cite{DUT2007,MIZ2009,BAL2009}. Those experimental values are not too far from the $T_2^{\ast}$ seen for Samples 3 and 4, where the nitrogen content was low enough that $^{13}$C is likely a major source of inhomogenous broadening due to dipolar coupling.

In conclusion, dipolar broadening due to paramagnetic nitrogen is predicted to be the dominant dephasing mechanism in the high-nitrogen-concentration samples, followed closely by dephasing due to other NV centers and $^{13}$C nuclear spins. For the low-nitrogen-concentration ([N]$\lesssim1~{\rm ppm}$) samples, dephasing due to the $^{13}$C nuclear spins was likely the dominant form of relaxation.

\subsection{Prospects for ultra-sensitive magnetometry}
\label{sec:prospects}
The performance of a magnetometer based on electronic spins is, in principle, limited by the spin-projection noise. The minimum detectable magnetic field, $\delta B$, is \cite{BUD2007}:
\begin{equation}
\label{eq:1}
\delta B \simeq \frac{1}{g_{s}\mu_B}\frac{1}{R\sqrt{\eta}}\frac{1}{\sqrt{N t T_2^{\ast}}},
\end{equation}
where  $R$ is the measurement contrast, $\eta$ is the detection efficiency, $N$ is the number of spin centers, and $t$ is the integration time. It is interesting to estimate the magnetic sensitivity of a sensor based under ideal conditions. We choose Sample 2, as it has the highest product of density and $T_2^{\ast}$, and we speculate that improved experimental techniques will allow the signal from nearly all NV centers to be collected with contrast $R\sim0.3$ and efficiency $\eta\sim0.1$. Using these values the quantum shot-noise limited sensitivity would be $\sim 150~{\rm fT/\sqrt{Hz}}$ for a $100~{\rm \mu m}$-scale magnetometer based on this sample. This approaches the sensitivity of superconducting quantum interference devices (SQUIDs) with similar spatial resolution \cite{CLA2004}.  Moreover, the isotropy of the system makes it easily size scalable, and the advantage of a such a compact device working in a broad range of temperatures--from room temperature down to liquid helium temperatures--makes NV-ensembles an attractive candidate for many types of magnetic-sensing applications. Of course, much work needs to be done to experimentally realize the shot-noise limit, including thorough analysis and mitigation of sources of decoherence and improved experimental techniques.  In particular, we note that the optimal optical pumping rate for this model magnetometer is on the order of $N_{NV}/T_2^{\ast}=2.8\times 10^{12}/(118~{\rm ns})$. This corresponds to approximately $8~{\rm W}$ for $532~{\rm nm}$ laser light. Thus, lengthening the effective relaxation time is essential in order to improve sensitivity while maintaining practical light powers.

\section{Conclusion}
Proposals for high-density ensemble magnetometers predict sensitivities rivaling the most sensitive atomic and superconducting magnetometers \cite{TAY2008}. Development of such magnetometers requires reliable production of diamond samples with control over the NV$^{\mbox{-}}$ concentration as well as other defects which may contribute to spin relaxation. In this work, ten diamond samples were prepared under very different conditions and their optical and spin-relaxation properties were characterized. The compiled results show that electron-irradiated samples annealed at $700^{\circ}$C for two hours can have NV$^{\mbox{-}}$ concentrations on the order of $10~{\rm ppm}$, a reasonable benchmark for $100~{\rm \mu m}$-scale magnetometry at the $100~\rm{fT/\sqrt{Hz}}$ sensitivity level. For samples with large initial nitrogen concentrations approaching $100~{\rm ppm}$, the highest irradiation dose used in this work, $\sim10^{19}~{\rm e^{\mbox{-}}/cm^{2}}$, was still insufficient for maximum nitrogen-to-NV$^{\mbox{-}}$ conversion. Nitrogen concentrations on the order of $50~{\rm ppm}$ remain in the samples, so in order to maximize the conversion efficiency, higher doses will be explored in future work. Conveniently, the irradiated samples remained mostly transparent in the spectral region of interest--absorption in the visible and near-infrared spectrum was due almost entirely to NV centers. Effective relaxation times above $100~{\rm ns}$ have been observed for some of the high-NV$^{\mbox{-}}$ concentration samples, likely limited by remaining paramagnetic nitrogen.  A $100~{\rm \mu m}$-sized magnetic sensor based on this sample operating under ideal conditions should achieve a sensitivity of $\sim150~\rm{fT/\sqrt{Hz}}$. In order to improve on this limit, future work is necessary to mitigate the dominant causes of relaxation.

\section{Acknowledgements}
The authors would like to thank B. Patton and S. Rochester for help with modeling and data visualization, T. Sauvage for proton irradiation, and P. Hemmer for extensive comments and suggestions. We also thank V. Bouchiat, C. Hovde, D. Twitchen, and R. Walsworth for providing helpful comments, R. Segalman and J. Sun for providing the Cary 50 spectrometer, and M. Donaldson and W. Vining for help with FTIR measurements. This work was supported by ONR-MURI and the Polish Ministry of Science (grant NN 505 0920 33).

\bibliographystyle{prsty}

\begin{thebibliography}{10}
\bibitem{SAN2009NOTE}
We say here ``effective'' energy splitting because in reality these vibrational
  levels are so strongly coupled to the continuum of delocalized phonons that
  there are no clear resonances.
\end{thebibliography}

\end{document}